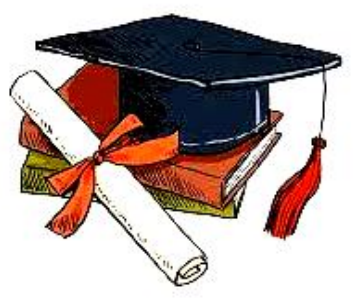

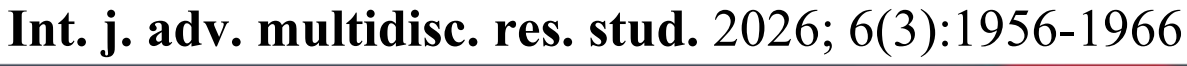

# Can AI-Assisted Inquiry Enhance Students' Decision-Making Skills in Socio-Scientific Issues? A Three-Group Experimental Study on Climate Change

**Dimitrios Gousopoulos**
Panteion University of Social and Political Sciences, Athens, Greece

Corresponding Author: **Dimitrios Gousopoulos**

**Abstract**
Climate change is a socio-scientific issue: it rests on science but cannot be settled by science, because any serious response forces people to weigh costs, values, and competing interests under uncertainty. Helping students make such decisions well is a central aim of science education, and the arrival of generative artificial intelligence raises a sharp question: does a conversational AI partner deepen students' reasoning, or simply do the thinking for them? This study tested whether AI-assisted inquiry improves secondary students' decision-making about climate change. Using a pretest-posttest design with three groups (AI-assisted inquiry, inquiry without AI, and traditional instruction; 270 students, 90 per group), reasoning was assessed across seven decision-making steps, from defining the problem to monitoring with adaptive management, using a four-level analytic rubric scored through content analysis with high inter-coder agreement. All three groups began at comparable, mostly low levels and all improved, but the gains differed sharply. The AI-assisted group improved most, ahead of inquiry-only and of traditional instruction. Between-group effect sizes on gains were large for AI-assisted versus traditional instruction and moderate-to-large for AI-assisted versus inquiry-only, with the clearest advantages on stakeholder engagement, alternatives, implementation, and monitoring. Within the AI group, the number of times students checked the AI's claims against the sources predicted their gains, and no student was flagged for over-reliance. The findings suggest that AI helps most when it is designed to question rather than to answer, and that the inquiry it is embedded in carries much of the benefit.

## Introduction

Many of the problems that matter most today are scientific and social at the same time. How a society should respond to climate change, manage a pandemic, regulate a new technology, or share limited natural resources are not questions that evidence can answer on its own. They ask people to balance trade-offs, take competing values and interests into account, live with uncertainty, and still reach a defensible decision [1, 2]. Problems of this kind are usually called socio-scientific issues, or SSI: open-ended, ill-structured, value-laden controversies that are tied to science conceptually or procedurally but are never reducible to it [3, 4]. Teaching students to reason about SSI and to make decisions on them has moved to the centre of science education, where it is treated as a core part of scientific literacy for citizenship [5, 6].

Climate change is the clearest example of the type. It is scientifically complex, plays out across long timescales and wide geographies, carries deep uncertainty, and is bound up with economics, social equity, and what we owe future generations [7, 8]. For adolescents it is also personal and emotionally loaded, yet their grasp of it tends to be patchy and shaped by the media, and many feel they can do little about it [9, 10]. International policy has responded by treating education as a key lever for climate action, and by calling for teaching that takes students past memorising facts toward informed, participatory decision-making [11, 12]. The job of the classroom, then, is not only to explain the science but to help young people deliberate and decide responsibly when things are genuinely uncertain.

Researchers rarely treat SSI decision-making as a single moment of choice. They model it as a process with several connected phases. Building on work in environmental and socio-scientific education, that process usually runs from defining the problem and its underlying causes, through collecting and making sense of evidence, identifying and engaging the people affected, generating possible courses of action, comparing those options against criteria such as feasibility, cost, and sustainability, planning how to act, and finally monitoring what happens and adjusting [13, 14, 15]. The same structure does double duty: it scaffolds the teaching, and it gives a clear set of dimensions for judging how well students reason at each stage.

The harder question is how to build these skills. Inquiry-based learning, where students investigate real questions, work with evidence, and construct and test explanations, has a solid track record for developing higher-order thinking, provided the inquiry is guided rather than left open [16, 17, 18]. The catch is that inquiry is demanding, and students usually need scaffolding: support that closes the gap between what they can manage alone and what they can manage with help, and that is slowly removed as they grow more capable [19, 20, 21]. For decades that support has come from teachers, classmates, printed prompts, or specialised software.

Generative AI, and large language models in particular, has introduced a scaffold of a different kind. A system like ChatGPT can hold an open-ended conversation, ask probing questions, offer explanations and counter-arguments, and give individual feedback on demand and at a scale no teacher can match [22, 23]. Early work hints that, when it is folded carefully into inquiry and constructivist settings, this kind of tool can lift engagement, support self-regulated learning, and deepen understanding [24, 25, 26, 27]. For a problem like climate change, where students have to assemble evidence, imagine other people's perspectives, and reason through trade-offs, a well-designed AI partner that keeps pushing them to justify their thinking might strengthen exactly the steps that good decisions depend on.

But the same tool can cut the other way. A growing literature warns that leaning on generative AI without structure can encourage cognitive offloading and what some call metacognitive laziness, where students take a fluent answer at face value and skip the effort that learning actually requires [28, 29, 30]. Studies of human-AI decision-making describe the same automation bias: people go along with the machine even when it is wrong, and only reflective prompts or deliberate friction restore a healthy, critical use of its advice [31, 32]. Large language models also produce confident text that is sometimes simply false, which makes uncritical acceptance more tempting and more costly [22]. So whether AI helps or hurts seems to depend less on the technology than on how students are asked to use it, and that puts the design of the task, and the comparison against inquiry without AI, right at the heart of the matter.

For all the attention the topic attracts, hard experimental evidence on whether AI-assisted inquiry improves students' SSI decision-making is thin. Most of what exists looks at higher education, centres on writing or coding rather than decision-making, or measures what students think of the tool rather than what they can actually do with it. Almost none of it pulls apart the contribution of the AI from the contribution of the inquiry it sits inside, which leaves a basic question unanswered: when students improve, is it the technology or the teaching approach doing the work?

This study is built to answer that question. It asks whether AI-assisted inquiry improves secondary students' decision-making about climate change, compared both with inquiry that uses no AI and with ordinary teacher-led instruction. By running three groups and scoring reasoning across the seven decision-making steps with a graded rubric, the design separates the effect of the AI from the effect of the inquiry and shows not just whether students improve but where and how. Three research questions guide the work: (RQ1) what level of decision-making skill about climate change do secondary students show at the outset, across the seven steps; (RQ2) does AI-assisted inquiry improve those skills more than inquiry without AI, and more than traditional instruction; and (RQ3) which steps improve most and least in each condition, and what do the patterns of within-level and between-level shifts tell us about how AI shapes students' reasoning?

**Socio-scientific issues and the aims of science education**

The idea of socio-scientific issues grew out of a wider shift in how science education defines its purpose. Roberts [6] drew a now-familiar line between Vision I, which puts canonical content and processes first, and Vision II, which starts from the situations citizens actually meet and asks what science they need to navigate them. SSI teaching belongs squarely to Vision II, and to a more recent Vision III that foregrounds the ethical, political, and relational sides of science and the goal of social transformation [34, 35]. On this view the point of learning science is not only to understand concepts but to take part, competently and responsibly, in public life where science is at stake. Dewey [33] made a version of this argument long before the term existed, insisting that genuine education grows out of reflective engagement with real, consequential problems.

In practice, SSI are described as contested, open-ended problems, such as genetic engineering, nuclear power, vaccination, and climate change, that have roots in science but cannot be resolved by science alone, because values, ethics, economics, and politics are always in play [1, 3]. Working with such issues has been shown to sharpen argumentation and moral reasoning, deepen students' understanding of how science works, and make them readier to consider more than one point of view [2, 36, 37, 38]. Reviews of the field trace its growth over two decades and its steady convergence on decision-making and reasoning as the abilities that matter most [4, 39].

**Decision-making and socio-scientific reasoning**

Deciding about an SSI draws on what Sadler [1] called informal reasoning: the work of generating and weighing positions on a messy, open problem. Looking at how students actually do this, Sadler and Zeidler [40] found three intertwined patterns, namely rationalistic, emotive, and intuitive, which is a useful reminder that a good SSI decision blends evidence with feeling and value, not just cold calculation. The broader ability to handle complexity, take other perspectives, pursue inquiry, and stay appropriately sceptical has been described as socio-scientific reasoning, a higher-order competence in its own right [36].

This kind of reasoning has deep roots in research on argument and judgment. Toulmin [41] gave us the basic anatomy of a sound argument, namely claim, evidence, and warrant, and Kuhn [42] and Kuhn and Udell [43] showed that the skills of argument develop gradually and respond to teaching, especially when students practise countering and rebutting. Work on argumentation in science classrooms has made much the same case [44, 45, 46, 47], and content knowledge turns out to matter for the quality of the argument students can build [48].

Environmental and science educators have turned these insights into structured models of decision-making. Eggert and Bogeholz [13] proposed a competence model in which students examine values and norms while they develop and judge possible solutions, and Gresch *et al.* [14] showed that teaching decision-making strategies explicitly raises secondary students' competence on SSI. Parallel frameworks for socio-environmental problem-solving lay out the same interlocking phases, namely defining the problem, gathering evidence, engaging stakeholders, generating and comparing options, implementing, and monitoring [15, 51]. Why structure helps is no mystery: Kahneman [49] and Baron [50] remind us that, left unsupported, people fall back on fast, intuitive, fragmentary judgment, a pattern that recurs in students' scientific reasoning when analytical processing is not deliberately engaged [52, 53]. The seven-step framework used in this study pulls these traditions together and serves at once as a teaching scaffold and an assessment grid.

**Inquiry-based learning and scaffolding**

Inquiry-based learning casts students as investigators who ask questions, work with data, and build and criticise explanations, much as scientists do [16, 55]. Dewey [33] and Kolb [54] supply the older pedigree here, in the conviction that we learn most from structured experience we then reflect on. Meta-analyses report that inquiry can markedly improve conceptual understanding and higher-order thinking, and that the gains are largest when students are properly guided [17, 56]. The well-known charge that minimally guided instruction fails [57] is best read not as a verdict against inquiry but as an argument for scaffolding it well [18].

Scaffolding traces back to Vygotsky's [58] zone of proximal development, the space between what a learner can do alone and what they can do with help from someone more capable. Wood, Bruner, and Ross [19] coined the scaffolding metaphor for the temporary, responsive support that fills that space and is taken away as the learner takes over. In inquiry, scaffolds can structure the task, complicate it on purpose to push students deeper, prompt them to reflect, or offer expert steers [20, 59, 60]. A synthesis of computer-based scaffolding studies finds reliably positive effects on learning [21]. Simulations and purpose-built environments such as WISE have long played this role [61, 62]; generative AI now offers a conversational version of the same idea, and how well it works will depend on how it is designed and used [63].

**Generative AI and large language models in education**

AI in education is not new; intelligent tutoring systems and learning analytics have been around for years [64, 65, 66, 67]. What changed with generative AI is the kind of interaction on offer. Where older systems followed fixed rules, large language models hold genuine conversations: they can pick up on confusion, explain the same idea several ways, and produce questions, hints, and feedback whenever a student asks [22, 23]. Reviews describe applications spreading quickly across levels and subjects, alongside steady worries about accuracy, fairness, assessment, and oversight [68, 69, 70, 71, 72], and uptake itself turns out to hinge on how teachers perceive and accept such tools [73].

In science and inquiry specifically, generative AI has been used to guide investigations, support self-regulated learning, give formative feedback, and build understanding, including custom GPT tutors designed around inquiry-based learning and theoretically grounded frameworks for upper-secondary work [26, 74], and good feedback is one of the strongest levers on learning we know of [75, 76, 77]. Ng *et al.* [24] found ChatGPT could help secondary students regulate their own science learning, and Xu *et al.* [25] showed that the benefit hinges on metacognitive support being built in. Evidence on conceptual gains is promising but mixed, and it depends heavily on teacher supervision and on how the task is set up [23, 78]. The upshot is that the AI works best not as an answer machine but as a dialogue partner that scaffolds, and the surrounding pedagogy decides which of those it becomes.

**The risks: Offloading, over-reliance, and epistemic agency**

Set against the promise are some well-documented hazards. Cognitive offloading, the use of external tools to lighten the mental load, can free up resources, but it can also starve the effortful processing that learning and memory feed on [28, 79]. Recent studies tie heavy use of AI tools to weaker critical thinking and to metacognitive laziness, particularly among younger users and when the tool is used without structure [29, 30]. Research on human-AI decision-making tells a similar story of automation bias: people accept AI suggestions even when they are wrong, and it takes deliberate friction, such as cognitive forcing functions and reflective prompts, to bring back appropriate, critical reliance [31, 32]. Because large language models sometimes state falsehoods fluently, the temptation to defer is real [22].

All of this bears on epistemic agency, students' sense that they, not the tool, are responsible for deciding what counts as knowledge in their own learning [80, 81]. An AI that hands over finished answers can quietly hollow that agency out; one that questions, challenges, and asks for justification can build it up. That is why AI literacy, knowing how to use, question, and ethically engage with these systems, is increasingly treated as a competence students need in its own right [82, 83, 84], and why some scholars urge caution rather than enthusiasm by default [85]. These arguments shaped the present study directly: the AI condition was engineered to prompt thinking rather than supply conclusions, and the non-AI inquiry group was included precisely so the comparison would be fair.

**Climate change as a socio-scientific issue in the classroom**

Climate change shows off both what makes SSI valuable to teach and what makes it hard. Responding to it means joining scientific evidence to social, economic, and ethical judgments that stretch across decades and continents [7, 8]. Reviews of climate change education point to what works, namely approaches that are personally relevant, deliberative, and oriented toward action, while naming the recurring obstacles: misconceptions, the sense that the problem is distant, and a low feeling of agency [7, 9, 10]. Traditions rooted in action competence and education for sustainable development press the same point, that the aim is to help learners deliberate and act, not merely to know [11, 86, 87].

Treated as an SSI, climate change asks for teaching that puts students in the position of actually deciding: naming the drivers, reading the data, taking the interests of different stakeholders seriously, weighing mitigation against adaptation, and planning and tracking what they would do. That made it an ideal setting to test whether AI-assisted inquiry can strengthen the structured reasoning that good decisions require, and it anchored this study's tasks, instruments, and rubric in a problem students recognise as their own.

## Materials and Methods

### Research Design

The study used a quasi-experimental pretest-posttest design with three parallel groups, set up so the contribution of the AI could be separated from the contribution of the inquiry to students' decision-making about climate change. The groups were: an AI-assisted inquiry group (AI-G), in which students investigated a climate change SSI with a generative-AI assistant configured to scaffold their reasoning; an inquiry-only group (IN-G), in which students ran the same investigation with conventional, non-AI scaffolds, namely printed prompts, curated resources, and teacher facilitation; and a traditional-instruction control group (C-G), which covered the same content through teacher-led instruction. Comparing AI-G with IN-G shows what AI adds on top of inquiry; comparing IN-G with C-G shows what inquiry itself is worth; and comparing AI-G with C-G captures the two combined.

All three groups completed the same structured worksheets before and after the intervention. The pre-test fixed a baseline; the post-test, given after the intervention, measured change. Teaching time was held equal across conditions, and decision-making was assessed individually in every group, so the comparison stayed fair at the level of the individual student even though the lessons differed in how much technology and collaboration they involved. In structure the design follows comparable pre-post SSI decision-making studies, extended here to three groups [13, 14].

### Participants

Participants were 270 secondary-school students drawn from public schools, assigned to the three conditions in equal numbers (90 per group) at the level of intact classes. They spanned lower and upper secondary (grades 8 to 11), with a mean age of 14.6 years (standard deviation 1.1); the sample was close to balanced by gender (133 male, 129 female, 8 reporting other or no response). Because whole classes rather than individuals were assigned to conditions, clustering at the class and school level is a feature of the design, and it is addressed both analytically and in the limitations.

The study followed institutional ethics requirements throughout. Parents gave informed consent and students gave assent before taking part, participation was voluntary, and all data were anonymised. The protocol was approved by the relevant ethics committee before any data were collected, and no identifying information was retained.

### The socio-scientific issue and learning context

The shared context was a climate change SSI pitched for adolescents: how a community should respond to worsening climate impacts through a mix of mitigation and adaptation, balancing the science against cost, fairness, and what is owed to future generations. The scenario was deliberately ill-structured and value-laden, with no single right answer, so that it exercised the full set of decision-making steps and the full range of reasoning, from the evidential to the moral [1, 2]. The core content, data, and case materials were identical across the three groups; only the way the lesson was taught and scaffolded changed, so any differences could be traced to the instruction rather than to the material.

### Intervention conditions

#### *AI-assisted inquiry group (AI-G)*

Students investigated the climate change SSI while working with a generative-AI assistant built on a large language model. The design took the offloading and over-reliance findings seriously [28, 29, 31]: through its system instructions and a guided prompting protocol, the assistant was set up to scaffold rather than to answer. It asked Socratic questions, pressed for evidence and justification, raised stakeholder perspectives and counter-arguments the student had missed, prompted comparison of options against stated criteria, and invited reflection, but it would not make the decision. Students received a short briefing on AI literacy first: how to prompt, how to check claims against the provided sources, and why a fluent answer is not automatically a correct one [82, 83, 88]. All AI use happened inside the supervised sessions, which protected fidelity and kept unsupervised use from muddying the results. Two process indicators were logged for this group only: the number of prompts a student sent, and the number of times they checked an AI claim against the provided sources.

#### *Inquiry-only group (IN-G)*

Students worked through the same investigation of the same SSI, in the same sequence of decision-making steps, but the conversational AI was replaced by conventional scaffolds: printed guiding questions matched as closely as possible to the AI protocol, curated resources and data, and teacher facilitation. This is well-designed guided inquiry [18, 59] without the generative AI, which is exactly what lets the AI's specific contribution show up when AI-G and IN-G are compared.

#### *Traditional-instruction control group (C-G)*

Students covered the same climate change content through teacher-led instruction grounded in constructivist practice, namely explanation, guided discussion, and small-group work, but without a student-driven investigation of the decision scenario and without generative AI. This mirrored ordinary classroom practice and served as the baseline both inquiry conditions were measured against. Teaching time was equalised across the three conditions, and students could reach the AI tool only during the AI-G sessions.

### Procedure

The study ran in three phases. First, in the pre-research phase, all groups completed the worksheets so that a baseline for decision-making about climate change was available. Second, in the intervention phase, each group experienced its own condition for an equal number of teaching hours. Third, in the post-research phase, students completed follow-up worksheets after a set interval; where feasible, a delayed post-test was added to check whether the skills last and transfer rather than simply spiking right after the lesson. Data collection ran over a single school term, which kept the setting realistic without losing control.

### Instruments

Decision-making was assessed with structured worksheets that every student completed individually, before and after the intervention. The worksheets were organised around the seven steps and asked students to define the climate change problem and its causes; gather and analyse relevant data; identify stakeholders and weigh their perspectives; generate alternative courses of action; compare those alternatives against criteria such as cost, effectiveness, environmental impact, and sustainability; plan how to implement a chosen course; and propose ways to monitor it and adapt. The scenarios were authentic and carried ethical and political weight, so they called on informal, moral, and socio-scientific reasoning rather than recall [1, 3].

Responses were scored with a four-level analytic rubric, namely Emerging, Developing, Proficient, and Advanced, scored 1 to 4 and defined separately for each of the seven steps. The rubric captured both how students reasoned through a decision and how deep and relevant the content of that reasoning was, building on established practice in designing critical-thinking and decision-making rubrics [89, 90]. Before the main study, science-education experts reviewed the instruments for content validity and the worksheets were piloted with a comparable group.

**Data analysis**

Analysis was mixed-methods. Students' worksheets were coded through systematic qualitative content analysis [91, 92, 93], guided by the rubric, within a wider mixed-methods frame [94, 95]. A calibration phase preceded full coding, and a subset of worksheets drawn proportionally from all three conditions was independently re-coded; agreement, reported as Cohen's kappa for each rubric dimension, was substantial to near-perfect and read against the usual benchmarks [96, 97]. For each student and each step, a shift score was computed as the post-intervention rubric level minus the pre-intervention level, capturing moves between levels (for example, Developing to Proficient) and, qualitatively, refinement within a level [98]. Baseline standing was summarised with the distribution of rubric levels across groups and steps, and baseline equivalence across the three groups was checked with a one-way analysis of variance on pre-test totals. Within-group change from pre to post was tested with paired-samples t-tests and summarised with paired effect sizes (Cohen's d and Hedges' g). Differences between groups in improvement were examined three ways: with one-way analysis of variance, and the Kruskal-Wallis test as a non-parametric check, on gain scores; with between-group effect sizes on gains; and with Pearson's chi-square on the post-test distribution of levels at each step, accompanied by Cramer's V, with a Bonferroni-type correction across the seven steps. Effect sizes were read against conventional thresholds [99, 100]. For the AI-assisted group, the two process indicators, namely prompts sent and evidence checks, were related to total gains with Pearson correlations. Because intact classes were assigned to conditions, clustering was treated as a real possibility and is addressed in the limitations.

**Results and Discussion**

**Baseline decision-making (RQ1)**

The three groups were statistically equivalent at baseline on total decision-making score (AI-G mean 12.3, IN-G mean 12.7, C-G mean 12.3; one-way analysis of variance $F(2, 267) = 0.32$, $p = 0.73$), so later differences cannot be attributed to a head start in any condition. Across the whole sample, students entered the study reasoning at a low level: pooling the groups, about 85% of responses sat at the Emerging or Developing levels across the seven steps, with only a small minority reaching Proficient or Advanced. The weakest steps were the most forward-looking ones, namely monitoring and adaptive management (mean 1.57) and decision implementation (mean 1.73), while problem identification was relatively the strongest (mean 1.99). In other words, students could begin to name a climate problem but struggled to turn a choice into a workable, trackable plan. Table 1 gives the full baseline picture.

**Table 1:** Baseline distribution of rubric levels across the seven decision-making steps (pooled sample, 270 students; % of students)

| Decision-making step | Emerging % | Developing % | Proficient % | Advanced % | Mean (1-4) |
|---|---|---|---|---|---|
| Problem identification | 32 | 44 | 18 | 6 | 1.99 |
| Data collection & analysis | 38 | 47 | 13 | 3 | 1.81 |
| Stakeholder engagement | 36 | 49 | 13 | 1 | 1.80 |
| Alternative solutions | 42 | 43 | 13 | 3 | 1.76 |
| Comparative evaluation | 42 | 40 | 16 | 2 | 1.77 |
| Decision implementation | 43 | 44 | 11 | 3 | 1.73 |
| Monitoring & adaptive mgmt | 52 | 41 | 6 | 1 | 1.57 |

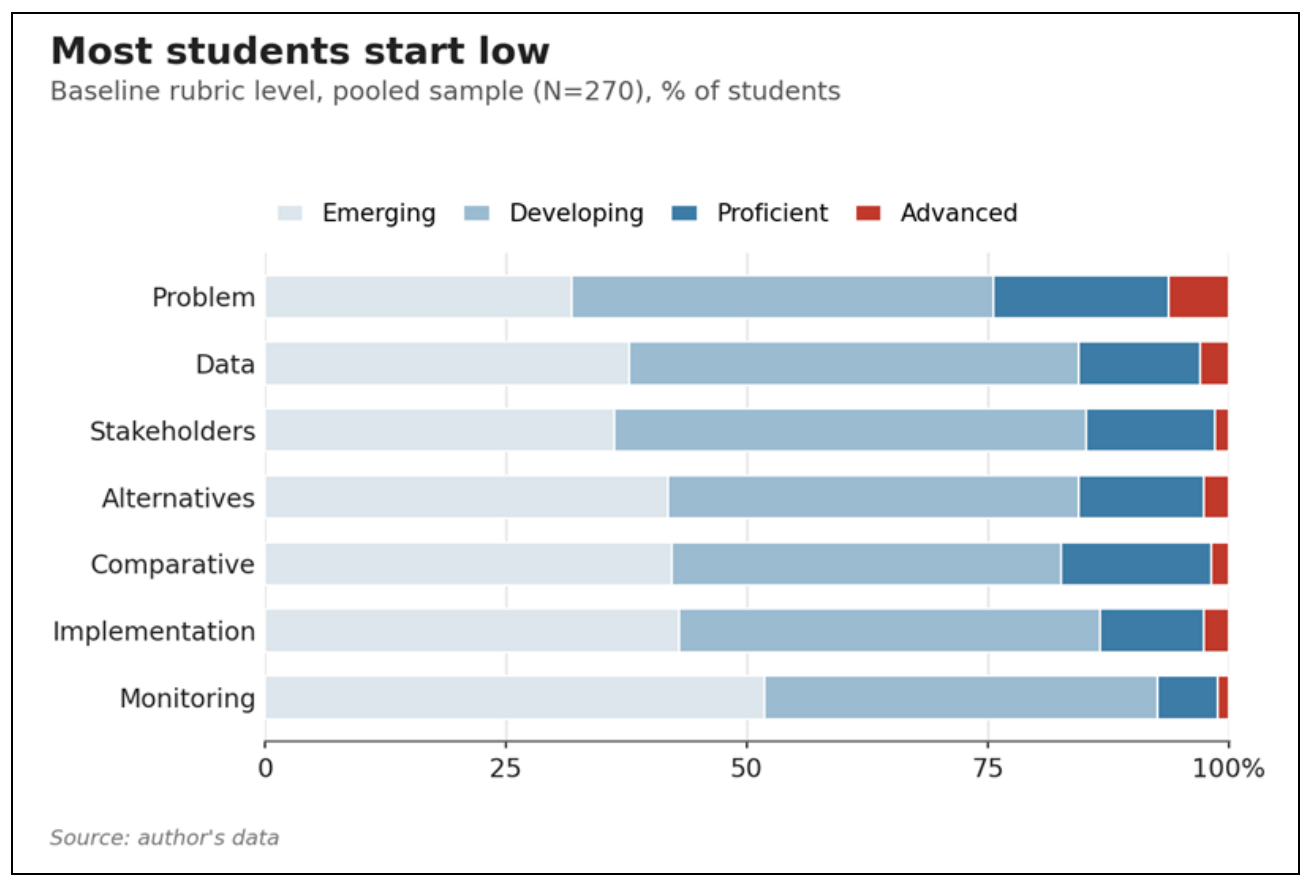

**Fig 1:** Baseline distribution of rubric levels by decision-making step (pooled sample, 270 students). Most students entered at the Emerging or Developing levels, especially on monitoring and implementation

**Did the groups improve differently? (RQ2)**

All three groups improved from pre-test to post-test, and every within-group change was statistically significant (all $p < 0.001$). But the size of the gain tracked the richness of the instruction. On the total score, the AI-assisted group moved from 12.3 to 18.3 (paired $d = 2.94$), the inquiry-only group from 12.7 to 17.3 ($d = 2.43$), and the traditional group from 12.3 to 14.9 ($d = 1.68$). The same ordering, AI-G ahead of IN-G and both well ahead of C-G, held for every one of the seven steps (Table 2). The largest AI-assisted gains appeared on monitoring and adaptive management (shift 1.07; $d = 1.37$) and on generating alternatives (shift 0.98; $d = 1.33$), the very steps where students had started weakest.

**Table 2:** Within-group pre and post means and paired effect size (Cohen's d) by step and group

| Step | AI-G pre | AI-G post | AI-G d | IN-G pre | IN-G post | IN-G d | C-G pre | C-G post | C-G d |
|---|---|---|---|---|---|---|---|---|---|
| Problem identification | 2.01 | 2.64 | 0.96 | 2.09 | 2.62 | 0.81 | 1.87 | 2.19 | 0.60 |
| Data collection & analysis | 1.76 | 2.44 | 0.99 | 1.89 | 2.48 | 0.96 | 1.78 | 2.11 | 0.62 |
| Stakeholder engagement | 1.81 | 2.70 | 1.42 | 1.81 | 2.48 | 1.05 | 1.78 | 2.12 | 0.69 |
| Alternative solutions | 1.70 | 2.68 | 1.33 | 1.76 | 2.54 | 1.11 | 1.83 | 2.24 | 0.74 |
| Comparative evaluation | 1.76 | 2.60 | 1.15 | 1.79 | 2.53 | 0.98 | 1.77 | 2.16 | 0.73 |
| Decision implementation | 1.72 | 2.61 | 1.56 | 1.79 | 2.41 | 0.96 | 1.68 | 2.07 | 0.65 |
| Monitoring & adaptive mgmt | 1.53 | 2.60 | 1.37 | 1.58 | 2.27 | 1.01 | 1.59 | 1.99 | 0.67 |
| Total (7-28) | 12.29 | 18.28 | 2.94 | 12.70 | 17.33 | 2.43 | 12.29 | 14.88 | 1.68 |

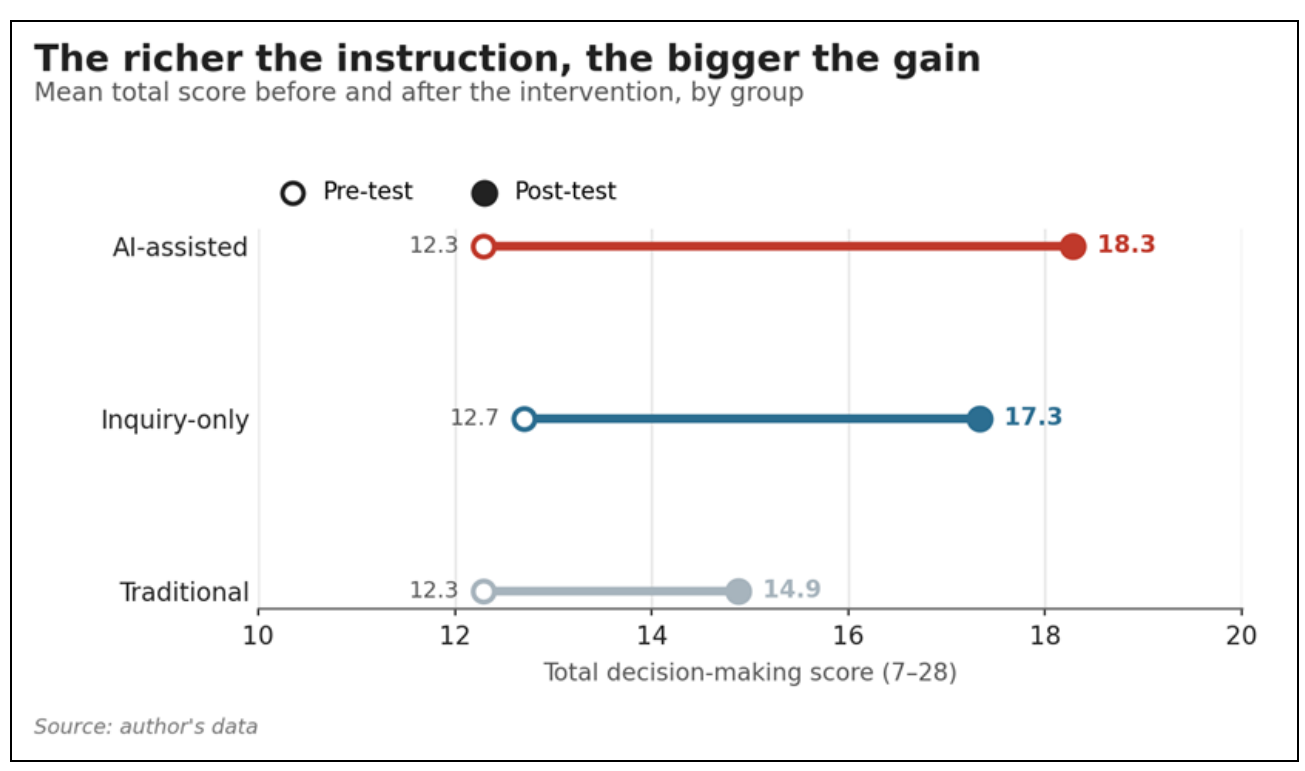

**Fig 2:** Mean total decision-making score before and after the intervention, by group. All groups started level; the AI-assisted group ended furthest ahead

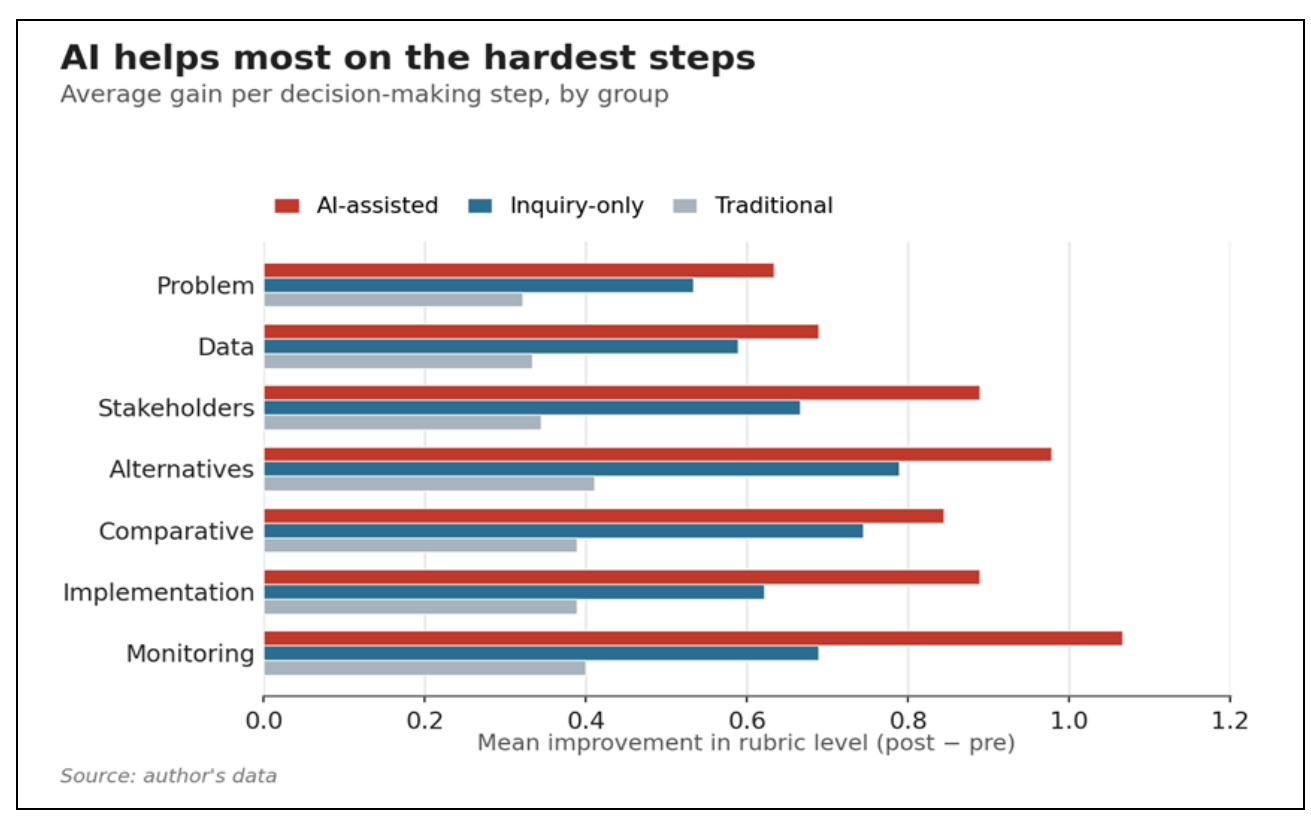

**Fig 3:** Average gain (post minus pre) on each decision-making step, by group. The AI-assisted group led on every step, most clearly on the later, more demanding ones

A one-way analysis of variance on gain scores confirmed that the three conditions differed for every step and for the total (total: $F(2, 267) = 77.9$, $p < 0.001$), with the Kruskal-Wallis test agreeing throughout. Pairwise tests on total gains separated all three groups cleanly: AI-G gained more than IN-G (5.99 versus 4.63; $t = 4.61$, $p < 0.001$), IN-G more than C-G (4.63 versus 2.59; $t = 7.91$, $p < 0.001$), and AI-G more than C-G ($t = 12.63$, $p < 0.001$). Translated into between-group effect sizes (Table 3), AI-assisted inquiry beat traditional instruction by a very large margin on the total score ($d = 1.88$), inquiry-only beat traditional instruction by a large margin ($d = 1.18$), and, in the key contrast, AI-assisted inquiry beat inquiry-only by a moderate-to-large margin ($d = 0.69$). The AI advantage over inquiry-only was widest on the later, more demanding steps: monitoring ($d = 0.52$), implementation ($d = 0.44$), and stakeholder engagement ($d = 0.35$).

**Table 3:** Between-group effect sizes (Cohen's d) on gain scores, by step

| Step | AI-G vs C-G | IN-G vs C-G | AI-G vs IN-G |
|---|---|---|---|
| Problem identification | 0.52 | 0.35 | 0.15 |
| Data collection & analysis | 0.57 | 0.44 | 0.15 |
| Stakeholder engagement | 0.96 | 0.56 | 0.35 |
| Alternative solutions | 0.87 | 0.59 | 0.26 |
| Comparative evaluation | 0.71 | 0.54 | 0.13 |
| Decision implementation | 0.86 | 0.38 | 0.44 |
| Monitoring & adaptive mgmt | 0.96 | 0.45 | 0.52 |
| Total | 1.88 | 1.18 | 0.69 |

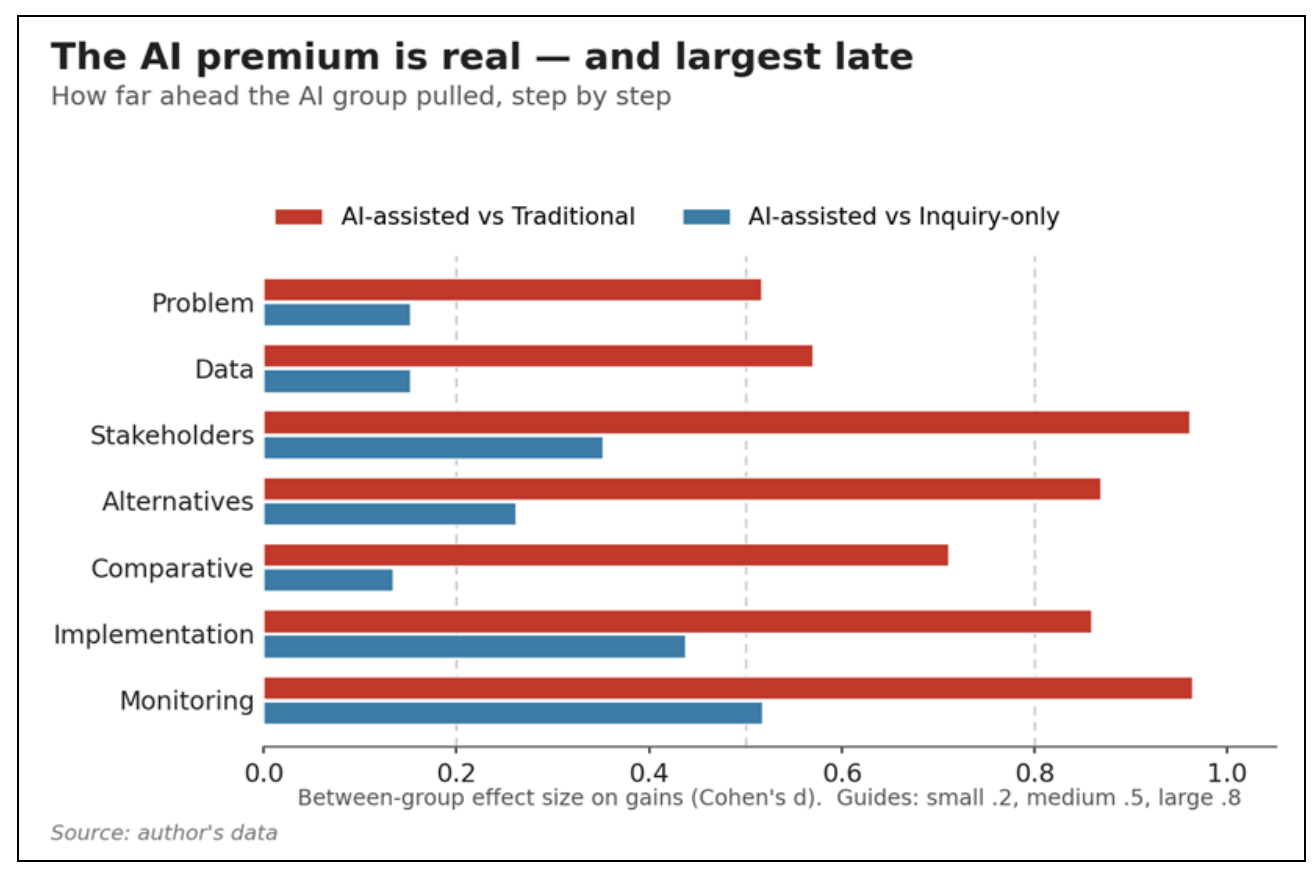

**Fig 4:** Between-group effect sizes (Cohen's d) on gains. The AI-assisted group's advantage over inquiry-only was widest on monitoring, implementation, and stakeholder engagement

### Where the groups ended up, and how they moved (RQ3)

Comparing the spread of post-test levels across the three groups with chi-square told the same story step by step (Table 4). After the Bonferroni-type correction (alpha = 0.05/7, about 0.007), the differences were clearest on stakeholder engagement, decision implementation, and monitoring and adaptive management, the steps that lean most on perspective-taking and forward planning. Data collection and analysis was the one step where the post-test distributions did not differ reliably between groups, suggesting that basic evidence-gathering improved fairly evenly regardless of condition. Effect sizes for these associations were small to moderate (Cramer's V from 0.13 to 0.19), as expected when comparing distributions rather than means.

**Table 4:** Pearson's chi-square comparing the post-test distribution of rubric levels across the three groups, by step (degrees of freedom = 6)

| Step | Chi-square | p | Cramer's V |
|---|---|---|---|
| Problem identification | 14.85 | 0.021 | 0.17 |
| Data collection & analysis | 8.51 | 0.203 | 0.13 |
| Stakeholder engagement | 20.18 | 0.003 | 0.19 |
| Alternative solutions | 13.84 | 0.032 | 0.16 |
| Comparative evaluation | 13.37 | 0.038 | 0.16 |
| Decision implementation | 17.92 | 0.006 | 0.18 |
| Monitoring & adaptive mgmt | 20.25 | 0.003 | 0.19 |

The shift analysis made the difference in movement vivid. Counting every student-by-step transition, 68% of the AI-assisted group's ratings moved up at least one rubric level, against 55% in the inquiry-only group and 33% in the traditional group; the rest held steady, and no student in any group regressed. Almost everyone improved on at least one step (AI-G 100%, IN-G 99%, C-G 91%), but the AI-assisted students were far more likely to make two- and three-level jumps, especially out of the Emerging band on the steps where they had started lowest. Within-level refinement, that is, qualitative growth that did not yet cross a level boundary, was common in all three groups and similar in volume, a reminder that progress was happening even where the rubric score stayed put. Table 5 summarises the movement.

**Table 5:** Movement across rubric levels (all student-by-step ratings pooled within group)

| Group | Stayed (n) | Transitioned up (n) | % transitioned | % improving at least 1 step |
|---|---|---|---|---|
| AI-G | 202 | 428 | 68% | 100% |
| IN-G | 284 | 346 | 55% | 99% |
| C-G | 419 | 211 | 33% | 91% |

**How the AI was used**

The process indicators logged for the AI-assisted group help explain the gains and speak directly to the over-reliance worry. On average students sent about 20 prompts (mean 19.8, standard deviation 4.1) and checked an AI claim against the provided sources roughly eight times (mean 8.2, standard deviation 2.2) over the intervention. Crucially, neither indicator pointed to passive consumption: the number of evidence checks correlated positively with total gains ($r = 0.52$, $p < 0.001$), as did the number of prompts ($r = 0.50$, $p < 0.001$). Students who interrogated and verified the AI more learned more, not less. Consistent with this, not one of the 90 AI-assisted students met the threshold for the over-reliance flag, which fits a design built to make the AI question rather than answer.

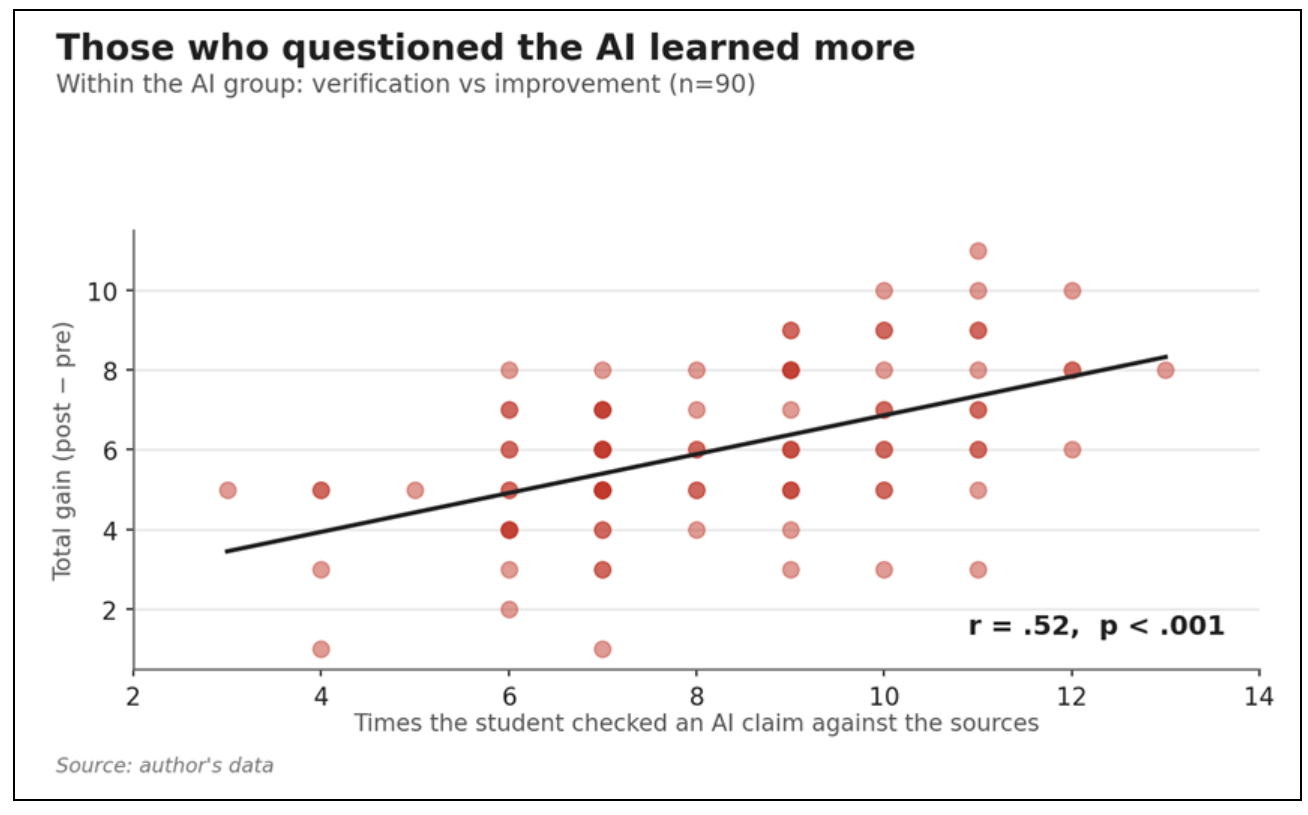

**Fig 5:** Within the AI-assisted group, students who checked the AI's claims against the sources more often gained more overall ($r = 0.52$, $p < 0.001$)

**Discussion**

This study asked a question that the rapid arrival of generative AI in classrooms has made urgent: does an AI partner deepen students' decision-making about a socio-scientific issue, or does it just do the thinking for them? On the evidence here, the answer is encouraging but conditional. Students who carried out a climate-change inquiry with a scaffolding AI improved their decision-making more than those who did the same inquiry without AI, and far more than those taught in the usual way. The three groups started level (RQ1) and ended clearly apart (RQ2), with the AI-assisted group making the largest and most numerous upward moves (RQ3).

The three-group design is what gives these gains their meaning. Because inquiry-only also beat traditional instruction by a wide margin, much of the benefit plainly comes from the inquiry itself, from putting students in the position of defining a problem, weighing evidence, and reasoning toward a defensible choice, exactly as the literature on guided inquiry would predict [17, 18]. But the AI added something on top of that, and not at random. Its advantage over inquiry-only was concentrated in the later, harder steps: engaging stakeholders, planning implementation, and monitoring with adaptation. These are the steps where a tireless conversational partner that keeps asking who else is affected, how one would know if it worked, and what one would change can do work that a printed prompt cannot, by following each student's reasoning and pushing on its specific weak points [20, 25].

The process data sharpen the interpretation. The students who gained most were the ones who prompted the AI more and, especially, who checked its claims against the sources more often, and no one tripped the over-reliance flag. This is the mirror image of the cognitive-offloading and automation-bias findings that warn of fluent answers accepted without thought [28, 29, 31]. It suggests those risks are not properties of the technology so much as of how it is used: when the tool is configured to question rather than to answer, and students are briefed to verify rather than to trust, engagement with the AI becomes effortful in the way learning requires. In that sense the result is less a verdict on AI than on a particular, deliberately constrained design of AI use, consistent with calls to treat AI literacy and critical reliance as part of the intervention, not an afterthought [82, 83].

That the weakest baseline steps, implementation and monitoring, were also where the AI helped most is worth dwelling on. Younger students routinely find it easier to name a problem than to turn a choice into a trackable plan, and ordinary instruction tends to leave that gap unaddressed [13, 14]. A scaffolding AI seems well suited to closing it, precisely because planning and monitoring benefit from being interrogated step by step. The one step that improved evenly across all conditions, basic data gathering, is also the one most schools already teach directly, which may explain why the extra scaffolding mattered less there.

For practice, the message is not to add AI but to design the AI to think with students rather than for them, and to keep it inside a genuine inquiry. A chatbot that hands over polished answers would likely produce the opposite of what was observed here; the gains rode on questioning, justification, and verification. Embedded that way, generative AI looks like a practical means of giving every student the kind of responsive, one-to-one prompting that a single teacher cannot provide to a full class, on exactly the higher-order steps that decision-making about climate change demands.

**Limitations**

Several limitations qualify these conclusions. Intact classes, not individual students, were assigned to conditions, so responses within a class or school may not be fully independent; although baseline equivalence was confirmed and the effects were large, clustering could inflate precision, and future work should model it explicitly or randomise at the class level. The assessment relied on written worksheets scored against a rubric; while inter-coder agreement was strong, worksheets capture reasoning imperfectly and may favour students who write fluently. The intervention was relatively short and the post-test followed soon after, so the durability and transfer of the gains remain open, and a delayed post-test would tell us whether the AI-assisted advantage lasts. The process indicators, though informative, are coarse, namely counts of prompts and checks rather than the content of the dialogue; richer logs and transcripts would reveal how, not just how much, students engaged the AI. Finally, the study was conducted with one age range on one socio-scientific issue in one educational system, so generalisation to other ages, issues, and contexts should be cautious.

## Conclusion

Set in the context of climate change, this study found that AI-assisted inquiry improved secondary students' decision-making more than the same inquiry without AI, and substantially more than traditional teaching, with the clearest advantages on the demanding work of engaging stakeholders and planning and monitoring action. The three-group design shows that much of the benefit belongs to inquiry, but that a carefully designed AI adds a real increment on top, and the process data show it does so without tipping students into passive over-reliance, since those who questioned and verified the AI most learned the most. The practical lesson is about design: generative AI helps students reason about socio-scientific issues when it is built to ask rather than to answer and embedded in authentic inquiry. Whether the advantage endures over time, transfers to other issues, and holds across ages and settings are the natural next questions, best pursued with larger, randomised samples and richer records of how students and AI actually talk to each other.